\documentclass[%
 aip,
cp,  
 amsmath,amssymb,
 reprint,%
]{revtex4-2}

\usepackage{graphicx}
\usepackage{dcolumn}
\usepackage{bm}

\usepackage[utf8]{inputenc}
\usepackage[T1]{fontenc}
\usepackage{mathptmx} 

\begin{document}

\title{Hybrid Stars with Hyperons and Strange Quark Matter}

\author{Wasif Husain} 
 \email[Corresponding author: ]{wasif.husain@adelaide.edu.au}
\author{Anthony W. Thomas}%
\affiliation{
  CSSM, Department of Physics, The University of Adelaide, Adelaide 5005, Australia.
}
\date{\today} 
\begin{abstract}
We consider the possibility of having hybrid stars with a phase transition from hadrons into strange matter at the core of a neutron star in $\beta$ equilibrium. For the hadron phase equation of state (EoS) the quark-meson coupling model is used, while the MIT bag model is used to describe the strange matter phase. The phase transition is treated using the Gibbs construction method and results are calculated and checked against the observational constaints imposed on the EoS. The results are also compared with the hadronic EoS including hyperons, F-QMC700.   
\end{abstract}
\maketitle

\section{\label{sec:level1}Introduction}
Neutron stars are astrophysical laboratories for the study of the behaviour of matter at high energy density. The interior of neutron stars, from the  atmosphere to the core, covers a wide range of densities from 1 gm/fm$^3$ to 10$^{15}$ gm/fm$^3$ or greater. The energy density 10$^{15}$ gm/fm$^3$ implies that the volume per baryon is not much larger than the size of a nucleon. This suggests that at high energy densities baryons tend to overlap, which is a strong indication that baryons may undergo a phase transition into deconfined quarks. Energetically 3-flavored quarks may be more stable than hadronic matter~\cite{Bodmer:1971we,Witten:1984rs} and recently the same was shown for 2-flavored quarks in Ref.~\cite{Holdom_2018}. 

Many calculations have been published \cite{Heiselberg:1999mq,Weber:2004kj,GLENDENNING2001393}, which consider the presence of hadrons, strange matter, and a hybrid phase. Glendenning was among the first to consider the conservation of charges in hadronic-quark hybrid stars and show that with a first order phase transition a mixture of phases may occur \cite{PhysRevD.46.1274}. The treatment of the transition phase in which conservation of global baryon number and global electric charge is ensured is called the Gibbs construction. The equilibrium condition needs the baryon chemical potential, the pressure and the temperature to be equal in transition phases within the subatomic particle mixture. The fraction of the volume for each substance present in the mixture can be calculated to satisfy the global constraints.

In this study, we investigate the effect of a possible phase transition from hadronic matter, including hyperons above a certain density, to strange quark matter at the core of the neutron star under strict constraints on the nuclear EoS. A comparison is made in order to ensure that the hybrid star reproduces tidal deformability measurements from neutron star mergers \cite{Alvarez_Castillo_2019,PhysRevC.98.045804,BURGIO201961,PhysRevLett.120.261103,PhysRevD.97.084038}. The model is constructed for cold neutron stars in $\beta$ equilibrium. In most studies \cite{Wu_2019} either the Gibbs or Maxwell construction is used for the description of the mixed phase. The Maxwell construction imposes local charge neutrality and is more complex than the Gibbs construction. The way by which phase transitions take place in nature, with local (Maxwell's construction) or global charge (Gibb's construction) neutrality, directly depends on the surface tension of the transition phases. As shown in Refs. \cite{PhysRevD.64.074017,Voskresensky_2003,Maruyama_2008,PhysRevC.86.025203,Lugones_2013,PhysRevC.88.025207,PhysRevC.95.015804} the surface tension for deconfinement phase transitions is model dependent.
 
For coexistence, the quark and hadronic phases must have equal baryon chemical potentials and pressures but different chemical potentials for the electron. The Gibbs and Maxwell constructions correspond to zero and very large surface tension values at the hadron-quark interface, respectively, which implies the mixed phase for the Gibbs construction has lower energy than the Maxwell construction \cite{Bhattacharyya_2010}.  In both the Gibbs and Maxwell constructions Coulomb and surface energies, like finite-size effects, are ignored. In the presence of surface and Coulomb energies, pasta structures are expected to occur in the hadron-quark mixed phase [16–21]. It is interesting to check whether hadron-quark mixed phase stars are consistent with the constraint imposed by neutron star mergers, particularly the tidal deformability.  
In this study, for the sake of simplicity, only the Gibbs construction is taken into account for the phase transition. 

\section{\label{sec:level2}Phase transition}
A phase transition from nuclear matter into hadronic matter can occur at higher energy density and we consider two possible phases. The first describes the hadron phase, while the second describes a quark deconfined phase. For the former, promising results were obtained in Ref.~\cite{Rikovska_Stone_2007}, using the quark meson coupling model 
(QMC model)~\cite{Guichon_2006,Guichon:1995ue,GUICHON1988235}. Indeed, that calculation was the first to show that one could indeed obtain neutron stars with masses as large as 2 $M_\odot$ even when hyperons were included, several years before such heavy stars were discovered. We use this model to define the hadron phase. To describe the strange matter that comes into existence after the phase transition, the Glendenning model has been adopted. The phase transtion is described by the Gibbs construction method.

\subsection{\label{sec:level3}Hadron Phase: QMC model}
The hadronic phase is defined by the QMC model~\cite{Rikovska_Stone_2007}, which is based upon the MIT bag model, with the structure self-consistently adjusting to the strong local scalar fields generated in desne matter. In Ref.~\cite{Rikovska_Stone_2007} the authors found a transition from nucleonic matter into hadronic matter, including hyperons, at an energy density around 450 MeV/fm$^3$ and the neutron star based on the corresponding EoS is consistent with observational contraints \cite{Rikovska_Stone_2007}.  

The parametrized EoS are given below (Eq.(1)). The parameterizations given in 
Ref.~\cite{Rikovska_Stone_2007} allow us to generate the EoS of N-QMC700 (pure nucleonic matter) and F-QMC700 (EoS that includes hyperons). For the energy density below 450 MeV/fm$^3$ N-QMC700 is used, while above 450 Mev/fm$^3$ up to the phase transition to strange quark matter F-QMC700 is taken into account. 
\begin{equation}
P = \frac{N_1\epsilon^{p1}}{1+e^{(\epsilon - r)/a}}  + \frac{N_2\epsilon^{p2}}{1+e^{-(\epsilon - r)/a}} .
\end{equation}
The constants $N_1$, $N_2$, $p_1$, $p_2$, '$r$' and '$a$' were presented in 
Ref.~ \cite{Rikovska_Stone_2007}. The fit is accurate up to the energy density 1200 MeV/fm$^3$ and for comparision purposes F-QMC700 is used up to that point.\\ The radius of the neutron star is very sensitive to the EoS at the crust and so, for the sake of accuracy of the radius, the Baym-Bethe-Pethick (BBP) EoS \cite{BAYM1971225} is employed for energy densities below 100 MeV/fm$^3$. 

\subsection{Quark Phase: Glendenning model}
There is much speculation suggesting that there may be a transition into strange matter in the high energy density region at the core of a star. In the study of Glendenning~\cite{GLENDENNING2001393}, the MIT bag model was used to describe the deconfined quark phase.

The EoS of a strange star matter (strange matter EoS or MIT EoS) is  then given by,  
P = $\frac{1}{3}(\epsilon$ - 4B),
where $P$ is the pressure, $\epsilon$ is the energy density and B is the bag constant, which is taken to be $10^{14}$gm/cm$^3$. This EoS has sometimes been used to describe stars made of strange matter entirely, while the crust of neutron star cannot be made of strange matter because at the crust the energy density is very low for nuclear matter to change into strange matter. Therefore, we use this equation at high energy densities and explore the phase transition from hadrons to deconfined quarks under the Gibbs constuction.

\subsection{Mixed Phase}
As the baryon number in both cases remains same, the nuclear EoS can be matched with the hadronic EoS at the point of transition. The transition from hadronic matter into strange matter is complex because in the transition baryon number and charge must remain conserved at the point of transition. The mixed phase, containing both hadronic and quark phases, has surface tension between the two interfaces along with a sharp interface. It has been suggested that the geometric structure of the mixed phase might change from droplets to rods, tubes and slabs with increasing baryon density. For the construction of the transition phase the Gibbs method is considered~\cite{Wu_2019}. 

\subsubsection{Gibbs Construction}
In this method, effects caused by  Coulomb  and surface tenstion contributions are ignored because only bulk contributions are considered. Thus the  mixed phase excludes any pasta-like structures. In this method, global charge neutrality is required, which suggests that both hadronic matter and quark matter can be charged separately. Therefore, the energy density of the mixed phase is given by
\begin{equation}
\epsilon_{MP} = u \epsilon_{QP} + (1 - u) \epsilon_{HP} \, .
\end{equation}
Equation (2) is the Gibbs equilibrium condition derived by minimizing the total energy density (excluding surface and Coulomb terms), where $\epsilon_{MP}$ is the energy density of the phase transition, $u$ is the volume fraction of quarks, $\epsilon_{QP}$ is the energy density of the quark phase and $\epsilon_{HP}$ is the energy density of hyperon phase.  The resulting equilibrium conditions~\cite{Wu_2019} for the phase transition are as follows  
\begin{equation}
P_{HP} = P_{QP}, \qquad \mu_u + \mu_e = \mu_d = \mu_s = \frac{1}{3}\mu_n +\frac{1}{3}\mu_e, \qquad \mu_p = \mu_n - \mu_e, \qquad \mu_\mu = \mu_e \, ,
\end{equation}
where $P_{HP}$ is the pressure of the hadron phase, $P_{QP}$ is the pressure of the quark phase, $\mu_u$, $\mu_e$, $\mu_d$, $\mu_s$, $\mu_n$ are the chemical potentials of the up quark, electron, down quark, strange quark and neutron, respectively. 

Equation (3) is to be solved for the phase transition. Equation (3) suggests that during the phase transition the pressure remains constant while global charge neutrality must be maintained and baryon number density must remain conserved. As the chemical potential increases more of the hadronic matter converts into strange matter. At a given baryon density, $n_b$, there are two independent chemical potentials, $\mu_n$ and
$\mu_e$, which can be determined by the constraints of global
charge neutrality and baryon number conservation. Solution of these equations will give the energy density of transition phase. Below this energy density there will be hadrons. At this energy density hadrons will start to transit into deconfined quarks, so we will have a mixed phase until they have completely converted into deconfined quarks in the core.

\subsection{Structural equations}
The structural equation for a static, spherical neutron star involves solving the TOV \cite{PhysRev.55.364}\cite{PhysRev.55.374}  equation, derived from the general theory of relativity, for a particular equation of state (EoS). The line element for such a compact star is given by (G = c = 1)
\begin{equation}
ds\textsuperscript{2} = -e^{2\Phi(r)}  dt^{2} + e^{2\Lambda(r)} dr^{2} + r^{2} d\theta^{2} +r^{2}\sin^{2}\theta d\phi^{2} \, .
\end{equation}
The TOV equation for the star is 
\begin{equation}
\frac{dP}{dr} = - \frac{[\epsilon(r)+P(r)][4\pi r^3P(r)+m(r)]}{r^2(1-\frac{2m(r)}{r})}, \qquad m(r) = 4\pi\int_{0}^{r}dr.r^2 \epsilon(r) \, ,
\end{equation}
where $m(r)$ is the energy inside radius $r$, $dr$ is the thickness of the mass shell, $r$ is the distance from the centre of the neutron star, $dP/dr$ is the pressure gradient and $\epsilon(r)$ is the energy density.
Equation (5) is to be integrated simultaneously with the EoS, from the centre to the surface of the neutron star to calculate the mass and the radius of the neutron star.

To determine the tidal Love number and tidal deformability, Hinderer and Flanagan's~\cite{Flanagan_2008,Hinderer_2008,Hinderer_2010} method is adopted. Suppose a spherically symmetric, static neutron star is placed in an external tidal quadrupole field ($\epsilon_{ij}$) and in response the neutron star develops a tidal quadrupole moment ($Q_{ij}$), then the star's tidal deformability is $\lambda$, where $Q_{ij}$ = $-\lambda \epsilon_{ij}$. Note also that  $\lambda = \frac{2}{3}k_2 R^5$,
where $k_2$ is the tidal Love number. To compute the tidal Love number, following the work of 
Ref.~\cite{Hinderer_2008,Hinderer_2010}, the perturbation metric of a spherically symmetric star in a tidal field in Regge Wheeler gauge \cite{Regge:1957td} is 
\begin{equation}
ds^2 = -e^{2\Phi(r)}[1 + H(r)Y_{20}(\theta,\phi)]dt^2 + e^{2\Lambda(r)}[1-H(r)Y_{20}(\theta,\phi)]dr^2 \\+ r^2[1-K(r)Y_{20}(\theta,\phi)](d\theta^2 + sin^2\theta d\phi^2) \, ,
\end{equation}
where H and $Y_{20}$ are the factors introduced by Regge-Wheeler gauge transformation (even parity perturbation) and $K$ is connected to $H$ by the relation $\frac{dK}{dr} = \frac{dH}{dr} + 2H\frac{d\Phi'}{dr}$. This leads to 
\begin{equation}
[-\frac{6e^{2\Lambda}}{r^2} - 2(\Phi')^2 + 2\Phi'' + \frac{3}{3}\Lambda' + \frac{7}{r}\Phi' - 2\Phi'\Lambda' + \frac{f}{r}(\Phi' + \Lambda')]H + [\frac{2}{r} + \Phi' - \Lambda']H' + H'' = 0 \, ,  
\end{equation}
where $f = \frac{d\epsilon}{dp}$. Equation (7) is to be integrated 
with the help of the TOV equations from just outside the center of the neutron star to the surface. For the sake of brevity, following the work given in Refs.~\cite{Hinderer_2008,Hinderer_2010}, we skip to the tidal Love number   
\begin{equation}
\begin{aligned}
k_2 = \frac{8C^5}{5}(1 - 2C)^2[2 + 2C(y-1) - y]\times [2C(6 - 3y + 3C(5y - 8)) + 4C^3(13 - 11y + C(3y -2) + 2C^2(1 + y)) + \\3(1 - 2C)^2(2 - y + 2C(y - 1))\times\log(1-2C)]^{-1} \, .
\end{aligned}
\end{equation}
Here $C = M/R$ and $y = R\beta (R)/H(R)$ ($\beta(r)$ = 2$a_0$r, H(r) = $a_0r^2$, r$\rightarrow$0 and $a_0$ is the expansion factor which is to be taken an arbitrary value). Once $k_2$ has been calculated then the tidal deformability ($\lambda$) can be calculated by the relation $\lambda = \frac{2}{3}k_2 R^5$.

\section{Results and discussion}
\begin{figure}[h!]
	\begin{center}
		\includegraphics[width=0.5\textwidth, height=0.3\textheight]{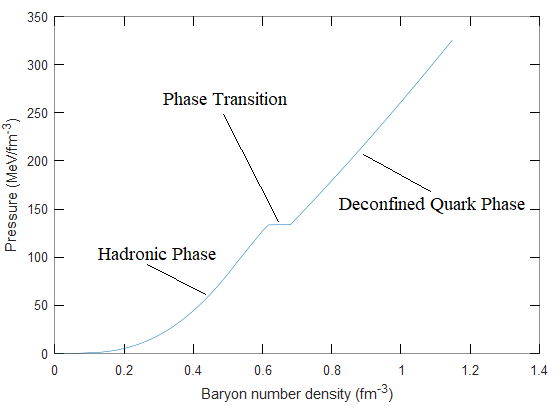}
		\caption{Pressure vs Baryon number density for phase transition.}
	\end{center}
\end{figure}
Figure 1 shows the pressure against baryon number density, including the phase transition. The flat part of the curve indicates the transition into strange matter. The baryon number density at which the hyperons start to transit into deconfined strange matter lies just above the point at which hyperons begin to appear, which is of order 0.45 fm$^3$.  Above the flat region, that is above the baryon number density 0.65 fm$^3$, we have strange matter (deconfined quarks). In a nutshell Fig. 1 shows, 
baryon number density $\leq$ 0.45 fm$^3$  $\rightarrow$ nucleonic matter,
0.45 fm$^3$ $\leq$ baryon number density $\leq$ 0.6 fm$^3$  $\rightarrow$ hyperonic matter,
0.6 fm$^3$ $\leq$ baryon number density $\leq$ 0.65 fm$^3$  $\rightarrow$ hyperon-strange matter mixed phase,
baryon number density $\geq$ 0.65 fm$^3$ $\rightarrow$ strange matter. 

Figure 2 shows the mass and the radius corresponding to the results given in Fig. 1. These were computed by solving the TOV equation~\cite{Tolman169}\cite{PhysRev.55.374}. The mass and the radius are consistent with the constraints imposed by gravitational wave measurements on the equation of state. That is, 
the radius of a neutron star of mass 1.4 solar mass, is less than 13 km and the hybrid star EoS predicts a  maximum neutron star mass above 2 solar mass \cite{2010Natur.467.1081D,Antoniadis1233232}. This is only slightly heavier than that predicted by the F-QMC700 EoS, which gave a maximum mass very close to (1.95$M_\odot$) 2 solar mass, before the measurement of 
Demorest {\it et al.}~\cite{2010Natur.467.1081D}.  
\begin{figure}[h!]
	\begin{center}
		\includegraphics[width=0.5\textwidth, height=0.25\textheight]{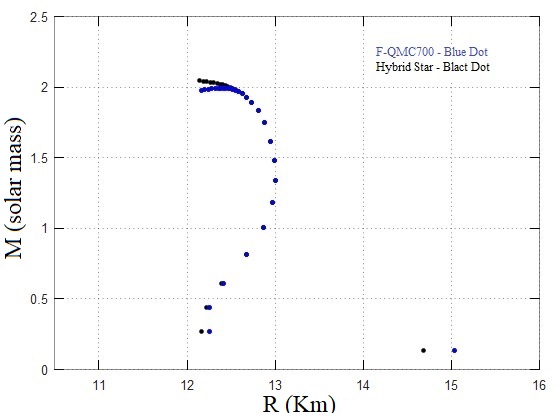}
		\caption{Mass vs Radius of neutron star for phase tansition EoS.}
	\end{center}
\end{figure}
\begin{figure}[h!]
	\begin{center}
		\includegraphics[width=0.5\textwidth, height=0.25\textheight]{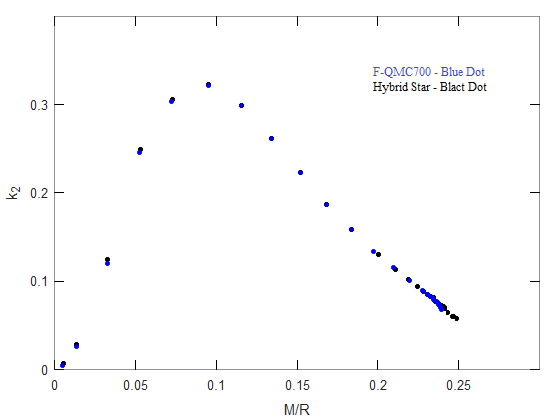}
		\caption{Tidal Love numver vs Compactness of the neutron stars for different EoSs.}
	\end{center}
\end{figure}

The tidal Love numbers calculated by the method suggested by \cite{Hinderer_2008,Hinderer_2010,Flanagan_2008} are shown in Fig.3, which compares \cite{Cardoso:2017cfl,Flanagan_2008} the relation between tidal Love numbers and compactness (M/R) of the neutron stars. The hybrid star EoS only differs slightly from F-QMC700 at the maximum neutron star compactness. But both EoSs follow the constraint imposed by the gravitational wave observations~\cite{Abbott_2017}. Therefore, one cannot exclude the possibility of a phase transition from hadrons to deconfined strange matter at the high energy density at the core of neutron stars. 

\section{Conclusion}
In this study, a hybrid star model was presented which was based on the quark-meson coupling model at low density and strange quark matter at high density. The model was shown to satisfy the constraints imposed by gravitational observations. The model predicts a maximum mass of over 2$M_\odot$ for a hybrid star with the radius less than 13 km~\cite{2010Natur.467.1081D,Antoniadis1233232}. 
The tidal Love number vs compactness plot, shown in Fig.3, for the hybrid star is also consistent with the neutron star merger observational constaint~\cite{Abbott_2017}. Thus, the EoS based on the QMC model at low densities allows a phase transition from hadrons to deconfined strange matter at higher energy densities at the core. 
There is therefore a possibilty of phase transition at the core of neutron stars which is perfectly consistent with the gravitational wave observational constraints.
In the future it will be interesting to investigate whether the how EoS  for hybrid stars is able to explain neutron star cooling, magnetic fields and glitches in the hybrid stars.  

\begin{acknowledgments}
W. Husain would like to thank Adelaide Scholarship International for financial support and  T.~Motta's valuable advice.
\end{acknowledgments}
\nocite{*}
\bibliography{HybridStar_APPC}
\end{document}